\documentclass[journal]{IEEEtran}
\ifCLASSINFOpdf
\else
\fi
\usepackage{acronym}
\usepackage{amsmath}
\usepackage{setspace}

\usepackage{graphicx} 
\graphicspath{{Figures/}} 

\usepackage{enumitem} 

\usepackage{lipsum} 

\usepackage{subfig} 

\usepackage{amsmath,amssymb,amsthm} 

\usepackage{varioref} 
\usepackage{url}
\usepackage[usenames, dvipsnames]{color}

\usepackage[final]{pdfpages}
\usepackage{pdflscape}
\usepackage{epsfig}
\usepackage{epstopdf}
\usepackage{multirow}
\usepackage{tabulary}
\usepackage{float}
\usepackage{tabularx}

\newcolumntype{Z}{>{\centering\let\newline\\\arraybackslash\hspace{0pt}}X}
\usepackage{subfig}
\usepackage{pbox}

\begin{document}
%
\title{Effect of Sensor Error on the Assessment of Seismic Building Damage}

\author{Ahmed Ibrahim,
	Ahmed Eltawil,
	Yunsu Na and
	Sherif El-Tawil
	\thanks{This work was supported by National Science Foundation under Award Numbers CMMI:1362547, CMMI:1362458 and OAC:1638186.
	}
	\thanks{A. Ibrahim and A. Eltawil are with the Electrical Engineering and Computer Science department, University of California, Irvine, CA 92697-2625 (email: \mbox{\{amibrah1,aeltawil\}@uci.edu})
		}
	\thanks{Y. Na and S. El-Tawil are with the Civil and Environmental Engineering department, University of Michigan, Ann Arbor, MI 48109-2125 (email: \mbox{\{yunsu,eltawil\}@umich.edu})
   	}
     \thanks{This work has been submitted to the IEEE for possible publication. Copyright may be transferred without notice, after which this version may no longer be accessible.}

}
	


%


\maketitle

\begin{abstract}
Natural disasters affect structural health of buildings, thus directly impacting public safety.
Continuous structural monitoring can be achieved by deploying an \ac{IoT} network of distributed sensors in buildings to capture floor movement. These sensors can be used to compute the displacements of each floor, which can then be employed to assess building damage after a seismic event. The peak relative floor displacement is computed, which is directly related to damage level according to government standards. With this information, the building inventory can be classified into immediate occupancy (IO), life safety (LS) or collapse prevention (CP) categories. In this work, we propose a \ac{ZUPT} technique to minimize displacement estimation error. Theoretical derivation and experimental validation are presented. In addition, we investigate modeling sensor error and \ac{IDR} distribution. Moreover, we discuss the impact of sensor error on the achieved building classification accuracy.

\end{abstract}


%
\IEEEpeerreviewmaketitle

\section{Introduction}
\acresetall
\acrodef{UCI}{University of California, Irvine}
Monitoring the structural health of buildings during and after natural disasters, such as earthquakes provides the public and policy makers with a clear view of the state of critical infrastructure that affects the safety and well being of the population.
Previous research on building damage assessment generally falls into one of two main categories: remote sensing techniques and sensor-based technology. In the former, optical images are captured using spacecraft or aircraft, then before and after image comparisons are performed to assess the damage. 
This technique is effective in detecting partial to complete collapse of buildings, however it cannot reliably detect incipient collapse because the resolution is too low \cite{image09}. On the other hand, sensor based technology uses an \ac{IoT} network of pre-installed sensors to capture the movement of a building during an event, enabling distributed, accurate and instantaneous monitoring of structures   
\cite{sensor06}.

\acrodef{ASCE}{American Society of Civil Engineers}
Measuring relative displacement of floors within a given building is used to calculate the \acp{IDR} for the building using (\ref{eqn:IDR}). \textcolor{black}{Documents released by government agencies and civil engineering societies such as \ac{FEMA} and \ac{ASCE} relate \ac{IDR} values to building damage level. Basically, these documents define two main critical thresholds of relative floor displacement of a given building, such that the building can be classified into one of three categories: \ac{IO}, \ac{LS} or \ac{CP}, which indicate that the building is either safe, needs further inspection or unsafe respectively. In other words, measuring the instantaneous relative floor displacement of a given building during an earthquake event is a good indicator of the structure state \cite{fema273,american2007seismic}.}  
\acrodef{IO}{immediate occupancy}
\acrodef{LS}{life safety}
\acrodef{CP}{collapse prevention}
\acrodef{IDR}{interstory drift ratio}
\begin{equation}
\label{eqn:IDR}
IDR=\frac{Displacement_{Floor}-Displacement_{NextFloor}}{FloorHeight}
\end{equation}

\acrodef{IoT}{internet of things}

Sensors are typically accelerometers, that are used to measure acceleration and consequently displacement. The cost of an accelerometer depends on several parameters such as dynamic range, linearity, bandwidth, output data rate, output noise and output type, i.e. analog or digital. Based on those specifications, cost ranges from a few dollars to a few thousand dollars.
According to \cite{cambridge_inertial}, sensor output noise is a major contributor to displacement measurement error, which is accentuated by double integration required to calculate displacement from acceleration. 
Hence, to minimize the system cost without sacrificing the accuracy, noise cancellation methodologies are adopted so that a cheaper less accurate device can be used instead of an expensive, highly accurate one.

\textcolor{black}{
While other technologies such as \ac{GPS} are widely used for localization and position estimation, accuracy becomes a major limiting factor in their suitability for structural health monitoring. As will be discussed later in the paper, to be useful, IDR values need to be estimated with an accuracy that is within a few centimeters from ground truth. This degree of accuracy is not possible using GPS alone, unless high-end GPS receiver is used, which is much more expensive than a standard GPS receiver \cite{GPS02,gpsGov}. Furthermore, GPS signals are not available indoors, which mandates outdoor installation for the sensing devices. Another approach to estimate position is to use vision based displacement estimation techniques as mentioned in \cite{cameras01}. Although this approach does not suffer from error accumulation, it faces other challenges such as, measurement error due to heat haze and ground motion, in addition to errors due to dim lighting and optical noise.}

\textcolor{black}{
Therefore, in this paper, we focused on studying the limits of using accelerometers to estimate structural displacement for a number of reasons: 1) earthquake event time is relatively short ($\sim$20-30 seconds) which results in bounded accumulated error that can be quantified, 2) accelerometers can work indoors which is not the case of \ac{GPS}, and 3) accelerometers are not affected by ambient light conditions as compared to cameras.}


\acrodef{GPS}{global positioning system}
\acrodef{ZUPT}{zero velocity update}

As mentioned earlier, accelerometer inherent noise is one of the main challenges in displacement estimation. However, noise cancellation can be achieved by depending on the fact that a disaster vibration intensity fades gradually and eventually stops at zero velocity and acceleration. In this case, the measured velocity at the \ac{EOS} reflects the accumulated error in the preceding samples, which can be used to minimize the estimation error. This technique is known as \ac{ZUPT} \cite{zupt10,zupt12}.

The main contributions of this work can be summarized as follows:
\begin{itemize}
	\item Derive how \ac{ZUPT} can be applied to minimize displacement estimation error; a theoretical derivation is presented and validated by shake table experiments.
	\item Study how displacement measurement error affects the accuracy of building damage classification based on its maximum \acp{IDR}.
	\item Present how different system parameters such as sensor noise and \ac{IDR} can be accurately modeled.
	\item Apply the derived methodology on a number of commercially available sensors to relate the probability of error versus duration of observation.
\end{itemize}
  The rest of the paper is organized as follows. In section \ref{sec:ZUPT}, \ac{ZUPT} algorithm is derived. Section \ref{sec:classification} describes the classification methodology and derives the probability of classification error, in addition to accelerometer noise and \ac{IDR} distribution modeling. System overall probability of error and sensor selection charts are presented in section \ref{sec:case_study}. Finally, the conclusions are drawn in section \ref{sec:conclusion}.

\section{Noise Cancellation}
\label{sec:ZUPT}
\begin{figure}
	\centering
	\includegraphics[width=1.05\linewidth]{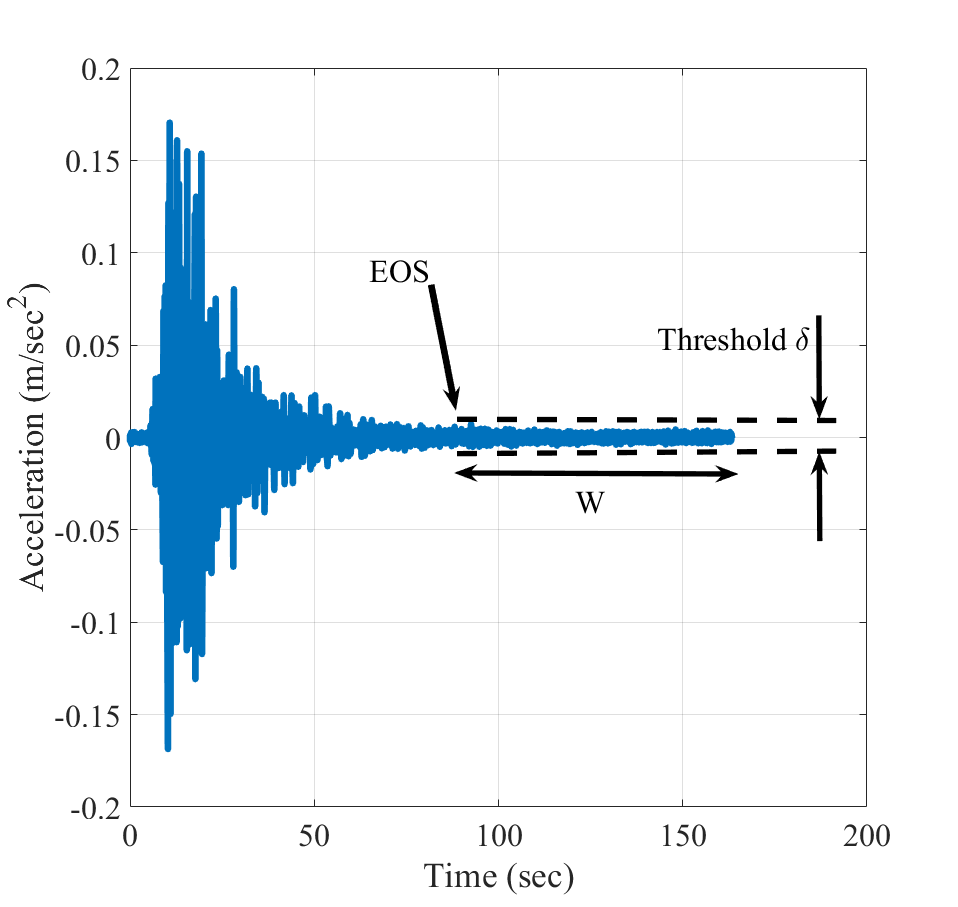}
	\caption{\ac{EOS} instant detection.}
	\label{fig:eos}
\end{figure}

An earthquake signal is characterized by stopping at zero acceleration and zero velocity. The \ac{EOS} instant can be detected when the absolute acceleration is below a certain threshold $\delta$ within a specified window of time $W$ as illustrated in figure \ref{fig:eos}. The selection of $W$ is arbitrary, whereas $\delta$ is dependent on the sensor noise. If the sensor noise \ac{STD} is $\sigma$, then we believe selecting $\delta = 3\sigma$ is a reasonable assumption, which indicates that the noise is below that threshold most of the time, In this region, the sensor has true zero velocity. Any non-zero velocity measured at this time is due to the sensor noise, and is correlated with the noise at shaking time. As mentioned before, using such characteristic in noise cancellation is known in the literature as \ac{ZUPT} \cite{zupt10,zupt12}.

\ac{ZUPT} has been used in inertial navigation systems, specifically pedestrian ones \cite{zupt11}. In such systems, navigation devices are mounted on a pedestrian's foot, which is known to be stationary on the ground once every step. The goal of applying \ac{ZUPT} in that case is to reset the velocity and prevent further error accumulation, which in turn reduces the error in upcoming velocity samples and consequently reduces the error in displacement estimation as well. However, in this work, we are only concerned in correcting displacement estimation for the time window \textbf{prior} to the \ac{EOS} instant, since that window contains the peak relative displacement which reflects the damage state.

\subsection{Noise cancellation using \ac {ZUPT} }
\label{sec:ZUPT_proof}
In this work, we are concerned with measuring floors' horizontal displacement. Hence, we assume that accelerometers will be oriented to measure only horizontal motion, i.e. gravity will not affect the reading. However, to account for miss-orientation, we consider a fraction $\alpha$ of the gravity will couple into the measurement. Assuming linear motion, $\alpha$ is constant throughout the motion, hence this constant can be removed with the sensor constant bias using long term averaging while the device is at rest. \textcolor{black}{In general, in case of curvilinear motion, $\alpha$ is not constant, and removing the gravity component in this case is more complex and can be addressed by using techniques described in \cite{gravity01}. Curvilinear motion is out of the scope of this paper and will be investigated in future work.}

True velocity is expressed by (\ref{eqn:ZUPT00}), where $a_{true}[k]$ is the ground truth $k^{th}$ horizontal acceleration sample. True displacement is obtained by (\ref{eqn:ZUPT02})-(\ref{eqn:ZUPT13}), where $A_{true}$ is the true acceleration vector and $P=[i,i-1,...1]^T$.

	\begin{align}
	v_{true}[i] 						=& \sum\limits_{k=1}^{i}  a_{true}[k]\  \Delta t										\label{eqn:ZUPT00}\\
	s_{true}[i]						=& \sum_{k=1}^{i}v_{true}[k] \Delta t=  \sum\limits_{k=1}^{i} \sum\limits_{j=1}^{k} a_{true}[j]\ \Delta t^2	\label{eqn:ZUPT02}
    \end{align}
    
	\begin{align}    
	=&  \sum\limits_{k=1}^{i} (i-k+1) a_{true}[k]\ \Delta t^2 										\label{eqn:ZUPT04}\\	
	=&  P^T A_{true}[1:i]	\Delta t^2																\label{eqn:ZUPT13}
	\end{align}

Measurement noise $z[k]$ is considered additive with zero mean, since constant bias is estimated by long term averaging and then subtracted from the measurement \cite{cambridge_inertial}. Hence, measured acceleration is expressed by (\ref{eqn:ZUPT15}). Consequently, measured displacement is shown by (\ref{eqn:ZUPT17})-(\ref{eqn:ZUPT19}). As a result, displacement error is shown by (\ref{eqn:ZUPT20})-(\ref{eqn:ZUPT06}), where $Z$ is the noise vector.

\begin{align}
a[i]						=& a_{true}[i] + z[i] \label{eqn:ZUPT15}\\
s[i]						=& \sum_{k=1}^{i}v[k] \Delta t=  \sum\limits_{k=1}^{i} \sum\limits_{j=1}^{k} a[j]\ \Delta t^2	\label{eqn:ZUPT17}\\
						=&   \sum\limits_{k=1}^{i} \sum\limits_{j=1}^{k} (a_{true}[j]+z[j])\ \Delta t^2	\label{eqn:ZUPT18}\\
						=&  s_{true}+ \sum\limits_{k=1}^{i} \sum\limits_{j=1}^{k} z[j]\ \Delta t^2	\label{eqn:ZUPT19}\\
e[i]						=&  \sum\limits_{k=1}^{i} \sum\limits_{j=1}^{k} z[j]\ \Delta t^2  										\label{eqn:ZUPT20}\\							
						=&  \sum\limits_{k=1}^{i} (i-k+1) z[k] \Delta t^2  										\label{eqn:ZUPT21}\\	
=&  P^T Z[1:i] \Delta t^2								  									\label{eqn:ZUPT06}
\end{align}

Let the shaking window length be $n$ samples, i.e., $v[n]$ is the measured velocity at the \ac{EOS} instant, which is equal to the accumulated noise since the true velocity at that instant is zero. As shown by (\ref{eqn:ZUPT07}), \ac{ZUPT} is applied at any given sample $i$, by multiplying $v(n)$ by certain coefficient $c_i$ and then adding the result to the displacement measurement. Thus, the modified displacement error is calculated by (\ref{eqn:ZUPT09}), where $Q$ is an $n\times 1$ vector of ones.

\begin{align}
s_{ZUPT}[i]					=&  P^T A[1:i] \Delta t^2 +c_i v[n]	\Delta t							  			\label{eqn:ZUPT07}\\			
e_{ZUPT}[i]					=&  P^T Z[1:i] \Delta t^2 +c_i \sum_{k=1}^{n} z[k] \Delta t^2								  			\nonumber\\			
=&  P^T Z[1:i] \Delta t^2 +c_iQ^T Z[1:n] \Delta t^2												\label{eqn:ZUPT09}
\end{align}
Let $\sigma_S^2[i]$ and $\sigma_{S\_ZUPT}^2[i]$ be the mean squared error in displacement at sample $i$ without and with applying \ac{ZUPT} respectively as shown by (\ref{eqn:di_error}) and (\ref{eqn:ZUPT10}).
\begin{align}
\sigma_{S}^2[i]=E[e^2]	=& P^T E[Z[1:i]Z[1:i]^T]P \Delta t^4\nonumber\\
=& P^T R_{ii}P \Delta t^4\label{eqn:di_error}\\
\sigma_{S\_ZUPT}^2[i]	=& E[e_{ZUPT}^2] \nonumber\\
=& P^T E[Z[1:i]Z[1:i]^T]P \Delta t^4\nonumber\\
&+c_i^2Q^T E[Z[1:n]Z[1:n]^T]Q \Delta t^4 \nonumber\\
&+2c_iP^TE[Z[1:i]Z[1:n]^T]Q \Delta t^4 \nonumber\\
=& P^T R_{ii}P \Delta t^4+c_i^2Q^T R_{nn} Q \Delta t^4 \nonumber\\
&+2c_i P^T R_{in} Q \Delta t^4 \label{eqn:ZUPT10}
\end{align}
where $E[.]$ denotes the expectation operator and the noise covariance $R_{in}$ and $R_{nn}$ are given by (\ref{eqn:R_in}) and (\ref{eqn:R_nn}) respectively.
The value of $c_i$ is calculated such that $\sigma_{S\_ZUPT}^2$ is minimized as shown by (\ref{eqn:ZUPT11}) and (\ref{eqn:ZUPT12}).

\begin{align}
R_{in}	   			&=
\begin{bmatrix}
\label{eqn:R_in}
r_{0}   	& r_{1}  	& \dots		&r_{i-1}		& \dots  	& r_{n-1}  		\\
r_{1}   	& r_{0}  	& \dots		&r_{i-2} 	& \dots  	& r_{n-2}  		\\
\vdots 		& \vdots 	& \ddots	&\vdots 	& \ddots 	& \vdots 		\\
r_{i-1}   	& r_{i-2}  	& \dots		&r_{0} 	& \dots  	& r_{n-i}
\end{bmatrix}
\end{align}

\begin{align}
R_{nn}	   			&=
\begin{bmatrix}
\label{eqn:R_nn}
r_{0}   	& r_{1}  		& \dots  	& r_{n-1}  		\\
r_{1}   	& r_{0}  		& \dots  	& r_{n-2}  	\\
\vdots 			& \vdots 			& \ddots 	& \vdots 			\\
r_{n-1}   	& r_{n-2}  	& \dots  	& r_{0}
\end{bmatrix}
\end{align}

\begin{align}
\frac{\partial \sigma_{S\_ZUPT}^2}{\partial c_i}=&0\label{eqn:ZUPT11}\\	
0							=& 2c_iQ^T R_{nn} Q +2P^T R_{in} Q  \nonumber\\
\Rightarrow	c_i				=& -\frac{P^T R_{in} Q}{Q^T R_{nn} Q} \label{eqn:ZUPT12}
\end{align}
Assuming that the noise can be modeled as a stationary process as will be illustrated in section \ref{sec:noise}, then $RQ\approx  Q\eta$, where $\eta$ is calculated by (\ref{eqn:eta}). Hence, equation (\ref{eqn:ZUPT12}) can be simplified as shown by (\ref{eqn:c_i2}). The resulting mean squared error $\sigma_{S\_ZUPT}^2$ is expressed by (\ref{eqn:result}).

\begin{align}
\eta					=& \sum_{k=-\infty}^{\infty} r_k \label{eqn:eta}\\
c_i 	=& -\frac{P^T Q[1:i]\eta}{Q^T Q\eta} \approx -\frac{i^2}{2n} \label{eqn:c_i2} 
\end{align}

\begin{align}
\sigma_{S\_ZUPT}^2[i] =& (P^T R_{ii} P + \frac{i^4}{4n^2} Q^T R_{nn} Q \nonumber\\
&- \frac{i^2}{n} P^T R_{in} Q)\Delta t^4  \label{eqn:result}
\end{align}

It is clear that the resulting mean squared error is a function of $R$ which depends on the noise characteristics. For example, in case of white noise, $R=\sigma^2 I$ and $\eta = \sigma^2$, where $I$ is the identity matrix and $\sigma^2$ is the noise variance. By substituting in (\ref{eqn:di_error}), the mean squared error without applying \ac{ZUPT} is calculated by (\ref{eqn:result_di_white0}) and can be simplified by (\ref{eqn:result_di_white4}) for sufficiently large $i$.

\begin{align}
\sigma_{S}^2[i] \bigg|_{white}			=& \sigma^2 P^T \ I \ P \Delta t^4 \label{eqn:result_di_white0}\\
									=& \sigma^2 P^T P \Delta t^4 \label{eqn:result_di_white1}\\
									=& \sigma^2 \Delta t^4 \sum_{k=1}^{i} k^2 \label{eqn:result_di_white2}\\
									=& \sigma^2 \Delta t^4 (\frac{i(i+1)(2i+1)}{6}) \label{eqn:result_di_white3}\\
									\approx& \sigma^2 \Delta t^4( \frac{i^3}{3}) \label{eqn:result_di_white4}
\end{align}

Similarly, by substituting in (\ref{eqn:result}), the mean squared error with applying \ac{ZUPT} is calculated by (\ref{eqn:result_white0}), where $G$ is an $i\times(n-i)$  matrix of zeros. For sufficiently large $i$, (\ref{eqn:result_white0}) can be simplified as shown by (\ref{eqn:result_white1})-(\ref{eqn:result_white5}).
 
\begin{align}
\sigma_{S\_ZUPT}^2[i] \bigg|_{white}	=& \sigma^2 (P^T\ I\ P + \frac{i^4}{4n^2} Q^T\ I\ Q \nonumber\\
									 &- \frac{i^2}{n} P^T [I\ G] Q) \Delta t^4 \label{eqn:result_white0}\\
									=& \sigma^2 (P^T P + \frac{i^4}{4n^2} Q^T\ Q \nonumber\\
									 &- \frac{i^2}{n} P^T Q[1:i]) \Delta t^4 \label{eqn:result_white1}\\
									=& \sigma^2 (\sum_{k=1}^{i} k^2 + \frac{i^4}{4n^2} \sum_{k=1}^{n} 1 \nonumber\\
									 &- \frac{i^2}{n} \sum_{k=1}^{i} k) \Delta t^4 \label{eqn:result_white2}\\
									=& \sigma^2 (\frac{i(i+1)(2i+1)}{6} + \frac{i^4}{4n}  \nonumber\\
									 &- \frac{i^2}{n} \frac{i(i+1)}{2}) \Delta t^4 \label{eqn:result_white3}\\	
									 \approx& \sigma^2 \Delta t^4( \frac{i^3}{3} +\frac{i^4}{4n} - \frac{i^4}{2n}) \label{eqn:result_white4}\\	
									 \approx& \sigma^2 \Delta t^4( \frac{i^3}{3} -\frac{i^4}{4n}) \label{eqn:result_white5}				
\end{align}

At the \ac{EOS} instant, i.e., at $i=n$, the mean squared error without and with applying \ac{ZUPT} are expressed by (\ref{eqn:result_di_white_eos}) and (\ref{eqn:result_white_eos}) respectively. Comparing both equations, it is concluded that using \ac{ZUPT} reduces the mean squared error by 75\% at the \ac{EOS} instant. For the rest of the paper we will refer to  $\sigma_{S\_ZUPT}$ as $\sigma_{S}$.

\begin{align}
\sigma_{S}^2 \bigg|_{white\ \&\ EOS } 	=&\sigma^2 \Delta t^4( \frac{n^3}{3}) \label{eqn:result_di_white_eos}	\\	
\sigma_{S\_ZUPT}^2 \bigg|_{white\ \&\ EOS }	=&\sigma^2 \Delta t^4( \frac{n^3}{12}) \label{eqn:result_white_eos}				
\end{align}



\acrodef{EOS}{end of shaking}

\acrodef{STD}{standard deviation}
\subsection{Experimental Validation}
\label{sec:exp_res}
\begin{figure}
	\centering
	\includegraphics[width=1.0\linewidth]{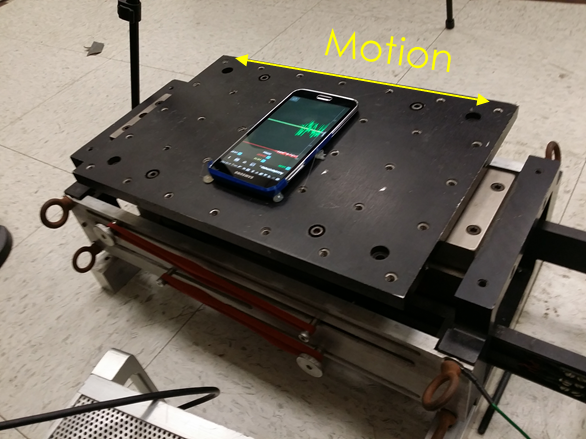}
	\caption{Shake table experiment setup.}
	\label{fig:experiment02}
\end{figure}

In order to validate the developed algorithm, shake table experiments have been performed. We have used different amplitudes of sinusoidal, triangular and random vibration profiles for a duration of 20 seconds. The sensing device is a smart phone that captures acceleration using its internal accelerometer and transmits the data to a PC. The phone internal accelerometer chip is Invensense MPU6500 which is a 6-axis inertial module that contains 3 accelerometers and 3 gyroscopes sensors, and is widely used in commercial devices \cite{MPU6500}. 
Figure \ref{fig:experiment02} shows the experimental setup.

\begin{figure}
	\centering
	\includegraphics[width=1.0\linewidth]{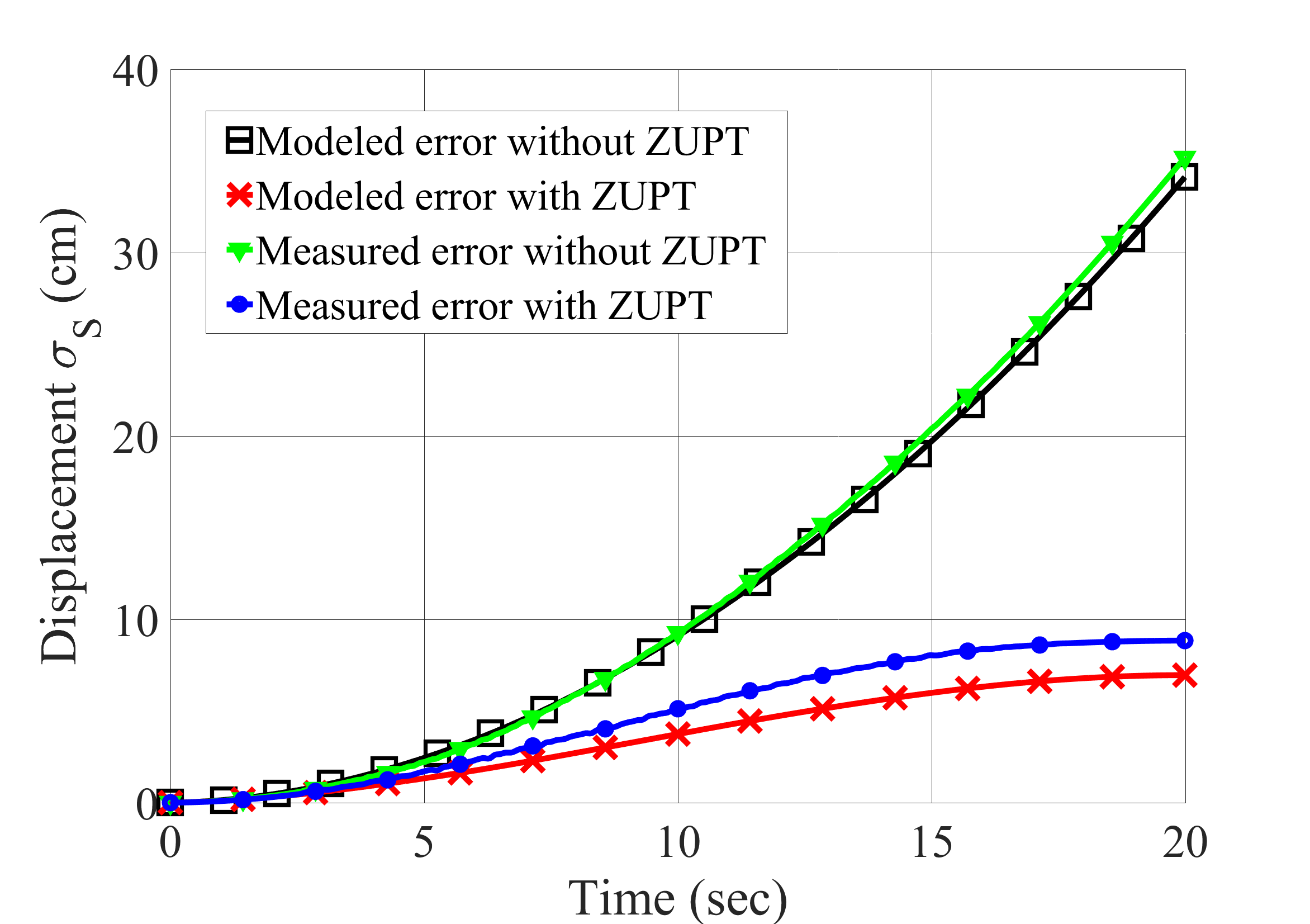}
	\caption{Experimental and theoretical displacement error with and without using \ac{ZUPT}. }
	\label{fig:exp_results}
\end{figure}

The motion starts and ends by zero velocity to mimic a seismic event. Figure \ref{fig:exp_results} shows the \ac{STD} of theoretical and measured error in displacement $\sigma_S$. It is clear that the measured error follows the theoretical one with and without applying the \ac {ZUPT} algorithm, and that applying \ac{ZUPT} decreases $\sigma_S$ by more than 75\%. It is worth noting that the reduction is greater than the one calculated in section \ref{sec:ZUPT_proof}, which is expected since in this experiment other noise sources were taken into account when modeling the sensor noise such as: \ac{BI} and \ac{RRW} rather than just white noise.

Measured error is slightly higher than the modeled one, due to the contribution of other sources of error, such as nonlinearity and sampling time jitter.

\acrodef{FEMA}{Federal Emergency Management Agency}

\section{Building Classification}
\label{sec:classification}
According to government documents, buildings are classified according to their damage state as \ac{IO}, \ac{LS} or \ac{CP} buildings. For instance, table  \ref{tab:IDR01} lists the IDR limits for steel moment frame buildings which are stated in \cite{fema273,american2007seismic}, and the corresponding physical tag used to signal the buildings' post-event condition. Hence, a building's performance can be assessed by comparing its peak \ac{IDR} to the predefined thresholds. Knowing the floor height, which is 4m in typical US construction, thresholds in \ac{IDR} corresponds to certain thresholds in relative floor displacement that we denote by $d0$ and $d1$. 

\begin{table}[]
	\centering
	\caption{Relation between \ac{IDR} and building state for steel moment frame buildings \cite{fema273,american2007seismic}.}
	\label{tab:IDR01}
	\begin{tabular}{|c|c|c|}
		\hline
		\textbf{IDR \%}         & \textbf{Building State}       & \textbf{Tag}          \\ \hline \hline
		$<0.7\%$                  & Immediate occupancy (IO)        & Green                 \\ \hline
		$0.7\%-5\%$               & Life safety (LS)              & Yellow                \\ \hline
		$>5\%$                  & Collapse prevention (CP)      & Red                   \\ \hline      		
	\end{tabular}
\end{table}

\acrodef{ARW}{angle random walk}
\acrodef{BI}{bias instability}
\acrodef{GM}{Gaussian Markov}
\acrodef{RRW}{rate random walk}
\acrodef{DRR}{drift rate ramp}

Let the true displacement of the two floors be denoted as $S_1$ and $S_2$, then the relative displacement $D$ is expressed by equation (\ref{eqn:rd0}). Since each displacement measurement has its own error, then the measured relative displacement $D_e$ is calculated by (\ref{eqn:rd1}), where $e_1$ and $e_2$ are the measurement error for $S_1$ and $S_2$ respectively.
\begin{align}
D &= S_2 - S_1		\label{eqn:rd0}\\
D_e &= S_2+e_2 - (S_1+e_1)		\label{eqn:rd1}\\
D_e &= S_2-S_1 + (e_2-e_1)		\label{eqn:rd2}
\end{align}

let $X = e_2-e_1$ then

\begin{align}
D_e &= D + X		\label{eqn:measure}
\end{align}
and knowing that the errors in both measurements are not correlated, then the mean squared error in relative displacement measurement is expressed by (\ref{eqn:sigmax0}). If identical sensors are used, then $\sigma_{S2}^2 = \sigma_{S1}^2 = \sigma_{S}^2$ and (\ref{eqn:sigmax0}) reduces to (\ref{eqn:sigmax1}).

\begin{align}
\sigma_X^2 	&= \sigma_{S2}^2 + \sigma_{S1}^2\label{eqn:sigmax0}\\
			&= 2\sigma_{S}^2 \label{eqn:sigmax1}
\end{align}

\begin{figure}
	\centering
	\includegraphics[width=1.0\linewidth]{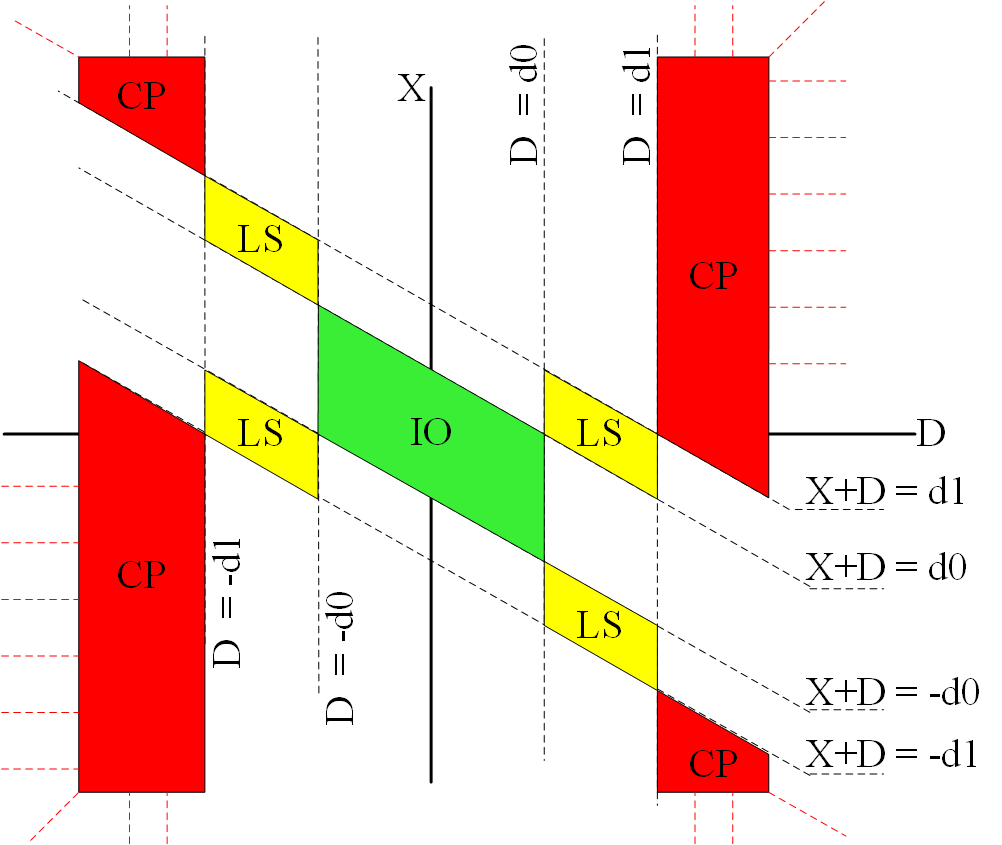}
	\caption{Integration regions for different $P(B|B_{true})$.}
	\label{fig:int_regions}
\end{figure}

\acrodef{PDF}{probability density function}
To evaluate classification accuracy, let $B$ and $B_{true}$ be the building's estimated and true states respectively. Equation (\ref{eqn:measure}) shows the measured relative displacement of two consecutive floors. The accuracy of the true classification of a building is obtained by evaluating the conditional probability $P(B|B_{true})$ as shown by (\ref{eqn:ptrue}).
\begin{align}
P(B|&B_{true}) = \frac{P(B\cap B_{true})}{P(B_{true})} \label{eqn:ptrue}
\end{align}
where $P(B \cap B_{true})$ and $P(B)$ are expressed by (\ref{eqn:joint_event}) and (\ref{eqn:event}).
\begin{align}
P(B \cap B_{true}) 	&= \int\int f_{X,D}(x,d)\ \textbf{d}d\ \textbf{d}x \label{eqn:joint_event}\\
P(B) 				&= \int f_{D}(d)\ \textbf{d}d \label{eqn:event}
\end{align}

where $f_{X,D}(x,d)$ is the joint \ac{PDF} of $X$ and $D$, and $f_D(d)$ is the marginal \ac{PDF} of $D$. The integral in (\ref{eqn:joint_event}) is done over the area shown in figure \ref{fig:int_regions}. Besides, limits of the integral in (\ref{eqn:event}) is given by table \ref{tab:limits}.

%

\begin{table}
	\centering
	\caption{Integration interval of equation (\ref{eqn:event}) for different $P(B|B_{true})$.}
	\label{tab:limits}
	\begin{tabularx}{1.0\linewidth}{|Z|Z|}
		\hline
		$P{(B|B_{true})}$                &Integration Region                 \\ \hline \hline
		$P{(B|IO)}$                     &$|D|<d_0$                          \\ \hline
		$P{(B|LS)}$                     &$d_0<|D|<d_1$                      \\ \hline
		$P{(B|CP)}$                     &$|D|>d_1$                      	\\ \hline
	\end{tabularx}
\end{table}

The measurement error only depends on the accelerometer itself and its inherent sources of noise, which is not related to the excitation signal. Hence, noise distribution is considered independent of \ac{IDR} distribution. As a result, the joint \ac{PDF} of $X$ and $D$ is expressed by (\ref{eqn:joint_pdf1}).
\begin{align}
f_{X,D}(x,d) =& f_X(x)f_D(d)  \label{eqn:joint_pdf1}
\end{align}
%
where $f_X(x)$ is the marginal \ac{PDF} of $X$. We will illustrate below how $f_X(x)$ and $f_D(d)$ can be modeled.

\subsection{Modeling}
\label{sec:modeling}
\subsubsection{Sensor Noise}
\label{sec:noise}
Acceleration measured by an accelerometer sensor is contaminated by several sources of noise, which can be modeled as: constant bias, \ac{ARW} (or velocity random walk), \ac{BI}, and \ac{RRW} (or acceleration random walk), where each of these is considered an independent Gaussian noise source with certain power spectral density. Since we are only concerned with relatively short durations, higher order noise sources such as \ac{DRR} are ignored and removed with the constant bias.

\acrodef{FIR}{finite impulse response}
According to \cite{noise01}, different noise sources can be modeled as white Gaussian noise shaped with a shaping \ac{FIR} function $H_f (z)$. Since the input of the \ac{FIR} filter is white Gaussian noise, i.e., \ac{WSS} noise process, then the generated noise is also \ac{WSS}. 

\acrodef{WSS}{wide sense stationary}

\begin{figure}
	\centering
	\includegraphics[width=1.07\linewidth]{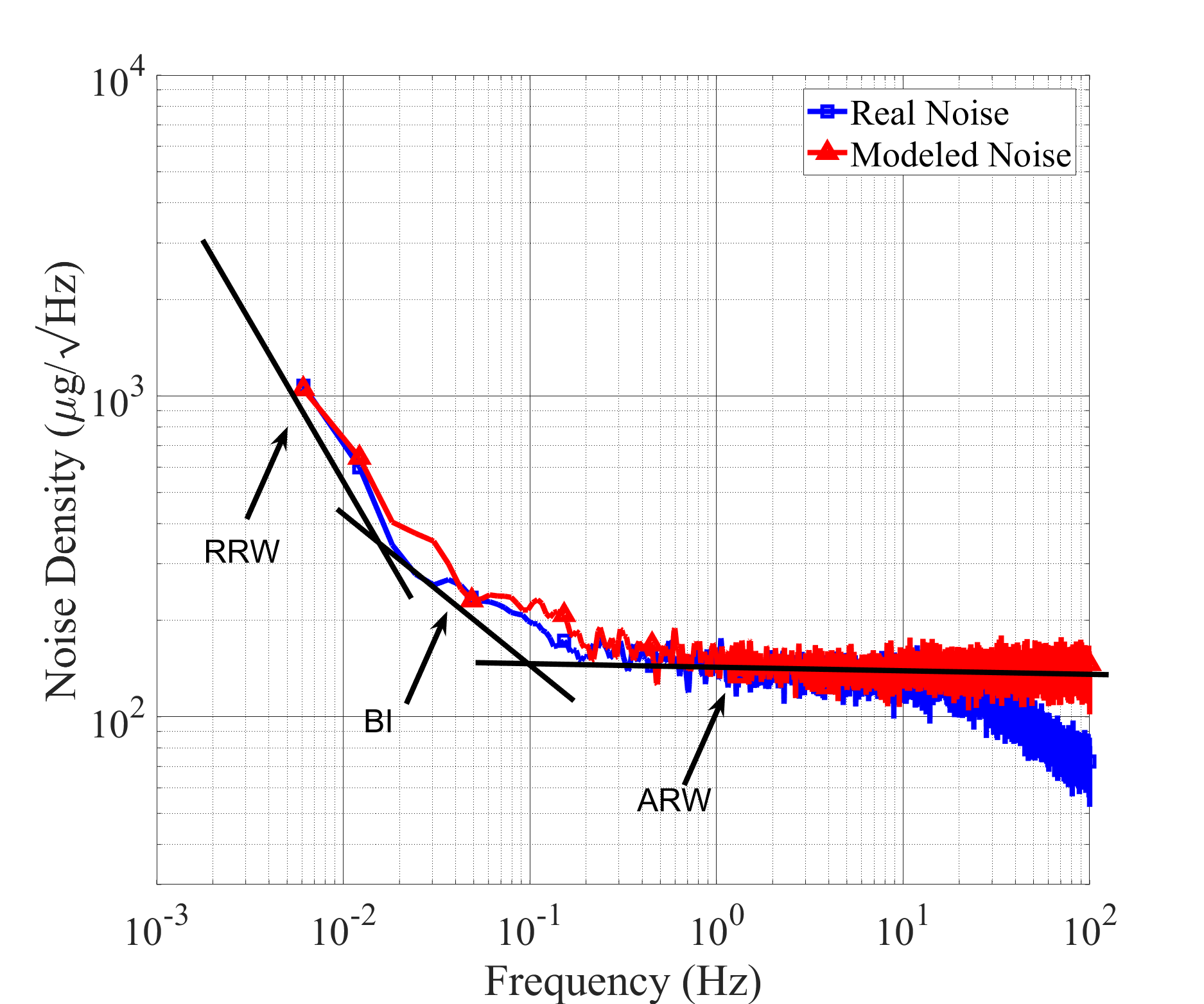}
	\caption{\ac{PSD} of accelerometer real and modeled noise.}
	\label{fig:PSD}
\end{figure}

\acrodef{PSD}{power spectral density}
\textit{Case Study}: For instance, consider the Invensense MPU6500, which is a 6-axis inertial module that utilizes 3 accelerometer and 3 gyroscope sensors. MPU6500 is widely used in commercial devices such as smartphones. To characterize the noise profile of the sensor, the output of the chip was recorded for 12 hours without motion. Using methods described in \cite{modeling03}, noise can be modeled as \ac {ARW}, \ac {BI} and \ac {RRW}, and the overall noise covariance matrix $R$ is calculated.
Figure \ref{fig:PSD} shows the real (measured) and modeled noise density of MPU6500 accelerometer sensor. It is clear that the noise model matches the real one at low frequency, whereas there is some discrepancy at high frequency, this is due to the fact that the sensor has a low pass filter in the output. However, since in our application the data is double integrated, the low frequency content is the main contributer to the displacement error. Therefore, the discrepancy at high frequency is irrelevant.

\begin{figure}
	\centering
	\includegraphics[width=1.07\linewidth]{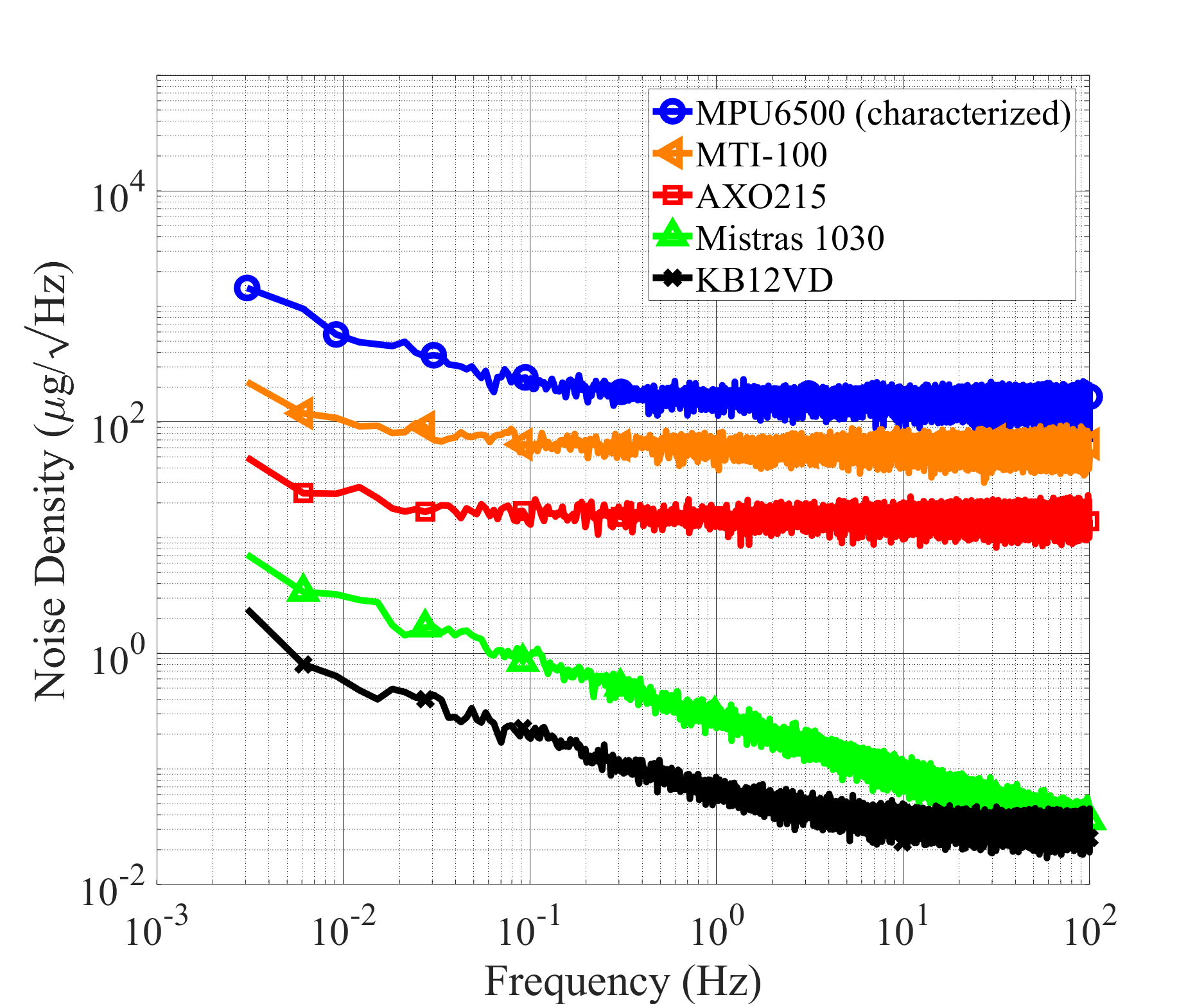}
	\caption{Noise density of several accelerometers based on their data sheets (except MPU6500 is measured and modeled).}
	\label{fig:PSDsensors}
\end{figure}

In order to provide a reference for comparison, we selected a number of sensors with different noise characteristics as summarized in table \ref{tab:sensors}. Figure \ref{fig:PSDsensors} shows the noise spectral density of the selected sensors based on their data sheets. From the figure, it is clear that some of the sensors noise can be approximated as white flat noise such as MTI100 and AXO215 sensors, whereas for other sensors higher order noise sources as \ac{BI} and \ac{ARW} should be considered.

\begin{table}
	\centering
	\caption{ Sensors noise characteristics according to their data sheets (except for MPU6500, noise is measured and characterized).}
	\label{tab:sensors}
	\begin{tabularx}{1.0\linewidth}{|Z|Z|Z|Z|}
		\hline
		\textbf{Sensor}     &\pbox{30cm}{\ \\ \textbf{Noise Density}\\ \centering{($\mu g /\sqrt{Hz}$)}}        &\pbox{30cm}{\  \textbf{In-run Bias} \\ \textbf{Stability} ($\mu g$)}     &\pbox{30cm}{\ \\ \textbf{Detailed PSD} \\ \centering{($\mu g /\sqrt{Hz}$)} }    \\ \hline \hline
		\pbox{30cm}{\  \\ MPU6500 \\ (measured)}               					& -                          	& -                     &\pbox{30cm}{\  \\ 700@0.01Hz \\ 200@0.1Hz \\ 150@10Hz}                  \\ \hline 
		MTI-100 \cite{MTI100}             					& 60                            & 15                    &-                          \\ \hline
		AXO215 \cite{AXO215}              					& 15                            & 3                     &-                          \\ \hline  
		Mistras1030 \cite{mistras1030}         					& -                         	& -                     &\pbox{30cm}{\ \\ 0.09@10Hz \\ 0.03@100Hz}                    \\ \hline  
		KB12VD \cite{KB12VD}              					& -                          	& -                     &\pbox{30cm}{\ \\ 0.3@0.1Hz \\ 0.06@1Hz \\ 0.03@10Hz}                  \\ \hline                   
	\end{tabularx}
\end{table}

\subsubsection{\ac{IDR} Distribution}

\begin{figure*}[htbp]
	\centering
	\subfloat[Building plan.]{\includegraphics[width=0.4\linewidth]{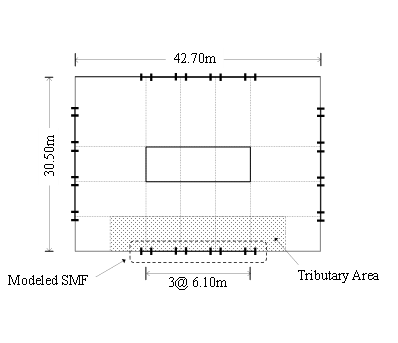}\label{fig:beam_a}}
	\subfloat[SMF model]{\includegraphics[width=0.6\linewidth]{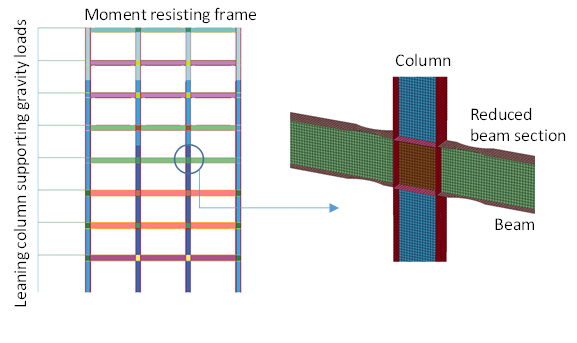}\label{fig:beam_bc}}
	\caption{Building plan and finite element modeling details of perimeter frame.}
\end{figure*}

\begin{figure*}[htbp]
	\centering
	\subfloat[50\% in 50 years hazard level]{\includegraphics[width=0.34\linewidth]{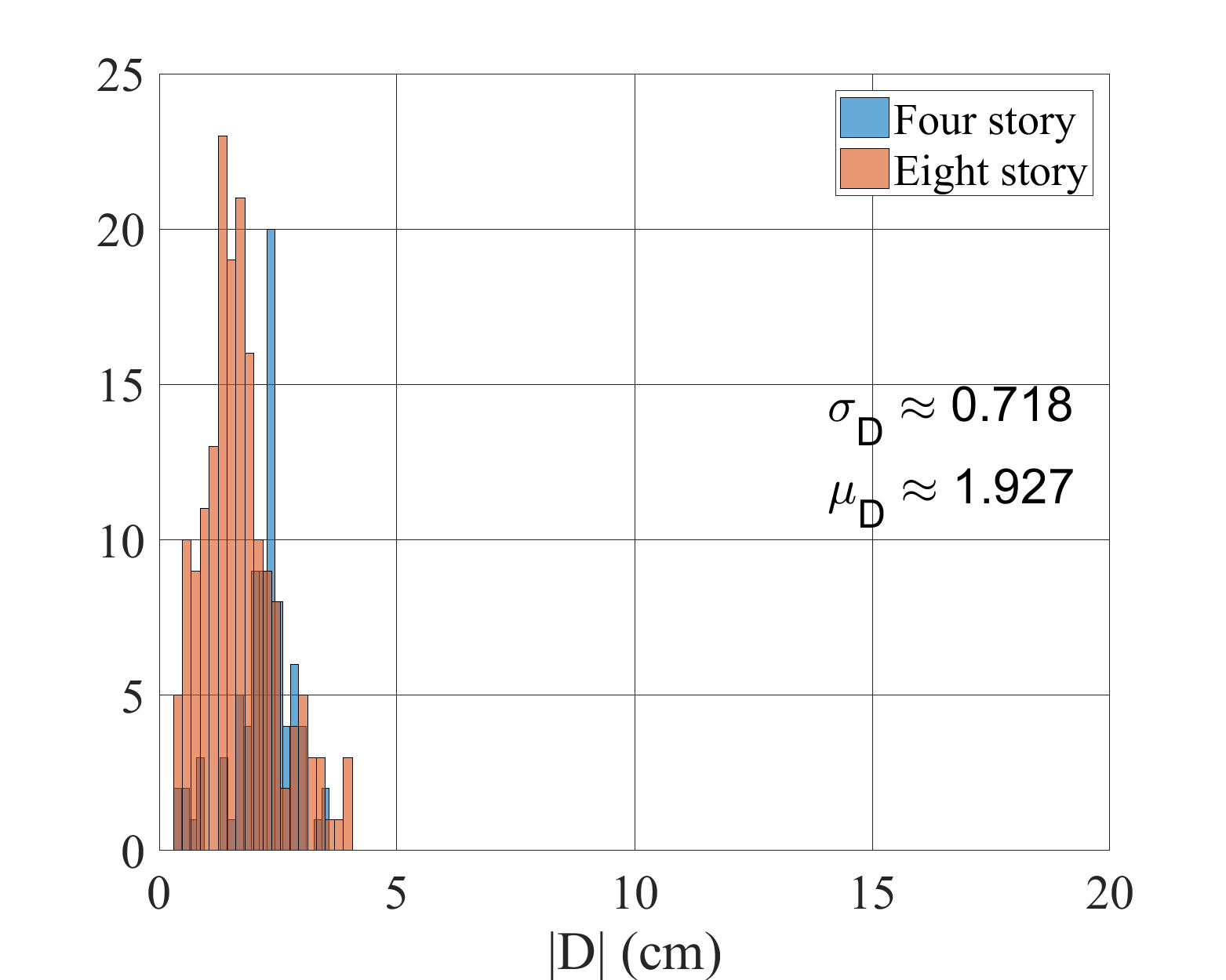}\label{fig:384storyhist}}
	\subfloat[10\% in 50 years hazard level]{\includegraphics[width=0.34\linewidth]{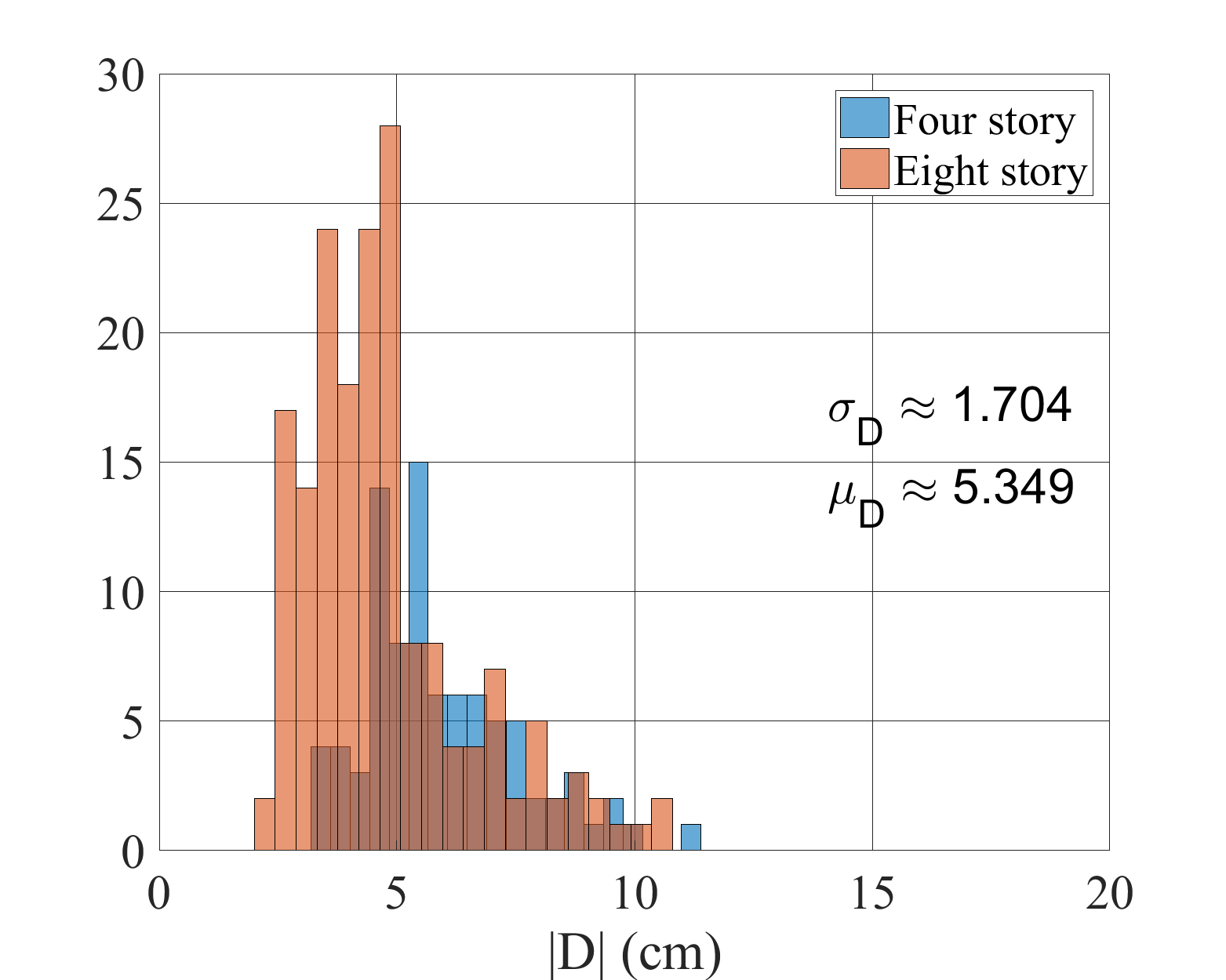}\label{fig:284storyhist}}
	\subfloat[2\% in 50 years hazard level]{\includegraphics[width=0.34\linewidth]{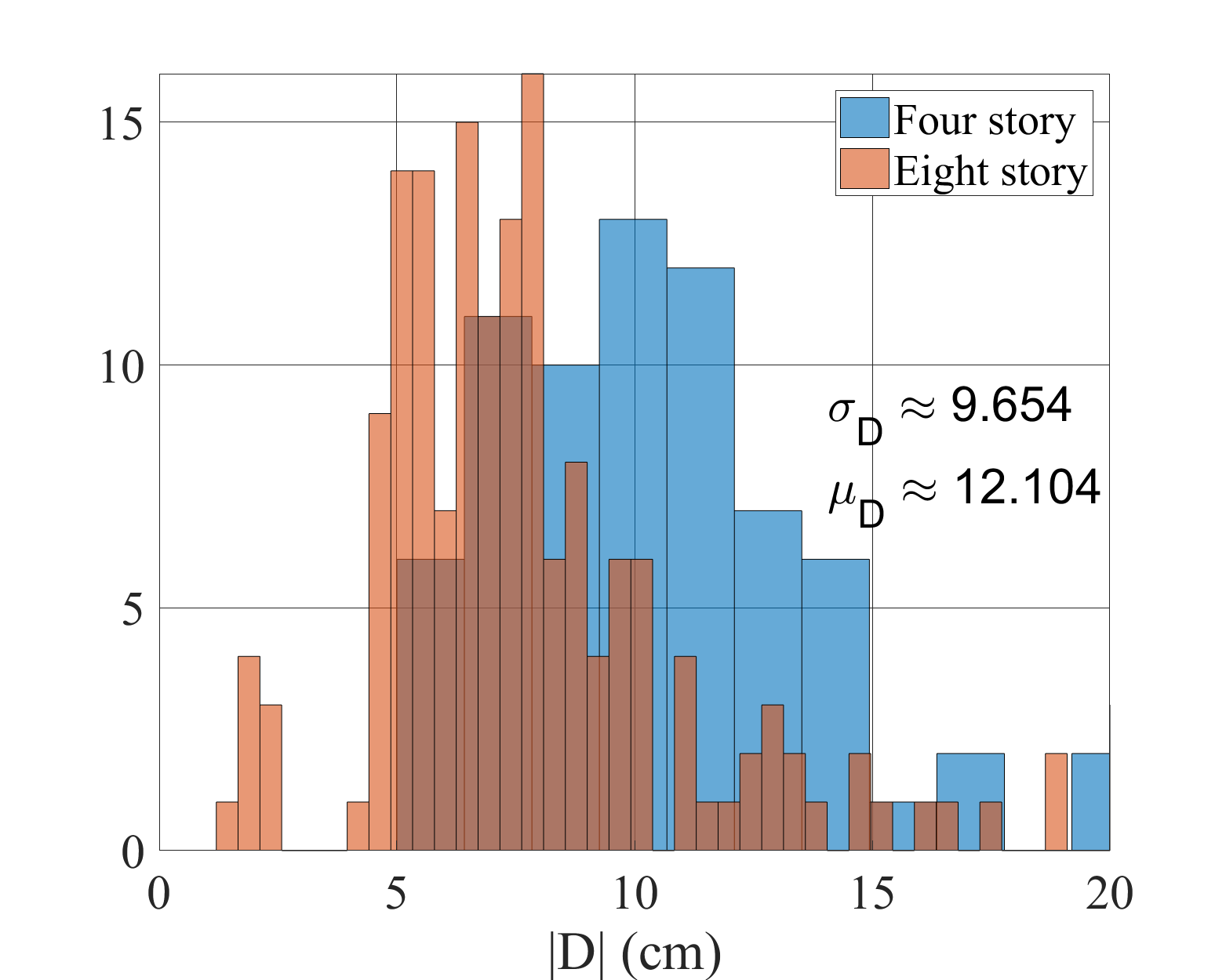}\label{fig:184storyhist}}
	\caption{Histogram of peak relative displacement of four and eight-story moment frame steel buildings.}
\end{figure*}

\label{sec:IDR_distribution}
\acrodef{USGS}{United States Geological Survey}
Simulation of building response is conducted in order to model the \ac{IDR} distribution as a result of earthquake excitation. We consider four and eight-story buildings designed by NIST \cite{nist} in Seattle to be representative of steel frame buildings.  The buildings have 42.7 m x 30.5 m plans as shown in figure \ref{fig:beam_a}. Three-bay perimeter steel special moment frames (SMFs) on each side of the building are used for the lateral load resisting system. The SMFs are designed with reduced beam sections (RBS). With respect to the type of soil, we consider site class $D$  which includes mixtures of dense clays, silts, and sands, which is the most common site class throughout the United States \cite{siteclass01}. The seismic design category is $D_{max}$, i.e., structures are expected to suffer from considerable rotational loads during strong earthquakes \cite{siteclass02}. As shown in figure \ref{fig:beam_bc}, finite element models of the SMFs are created using HyperMesh \cite{hypermesh} and analyzed using the commercial code LS-DYNA \cite{livermore}. The steel is ASTM-A992 and its engineering stress-strain properties are converted into true stress-strain data then assigned to the finite elements as done in \cite{momentframe01}. Gravity loads from the tributary area shown in figure \ref{fig:beam_a} are directly applied to the frame and the remainder of the gravity loads are applied to a leaning column connected to the SMF by truss members. Mass weighted damping of 2.5\% is assumed at the first mode period of the SMFs. Additional modeling details can be found in \cite{momentframe01}.

The distributions of peak relative displacement are computed for three seismic hazard levels: 2\% probability of exceedance in 50 years, 10\% in 50 years, and 50\% in 50 years. Eleven seismic records are selected from the Far-Field ground motion record set in FEMA \cite{fema695} and scaled to the three specified hazard levels at the first period of each building, resulting in 33 records for each building. The first period spectral accelerations corresponding to the three hazard levels are 0.55g, 0.26g, and 0.07g for the four-story building and 0.41g, 0.17g, and 0.04g for the eight-story building, respectively. Each building is then subjected to the scaled seismic records for each hazard level and the peak relative displacement is computed. The histogram of peak relative displacement is shown in figures \ref{fig:384storyhist} through \ref{fig:184storyhist}. The distribution can be approximated as Gaussian with mean $\mu_D$ and variance $\sigma_D^2$ that depend on the hazard level, with slight variation depending on the building type.

\subsubsection{Earthquake Strong Motion Duration}
\begin{figure}
	\centering
	\includegraphics[width=1.0\linewidth]{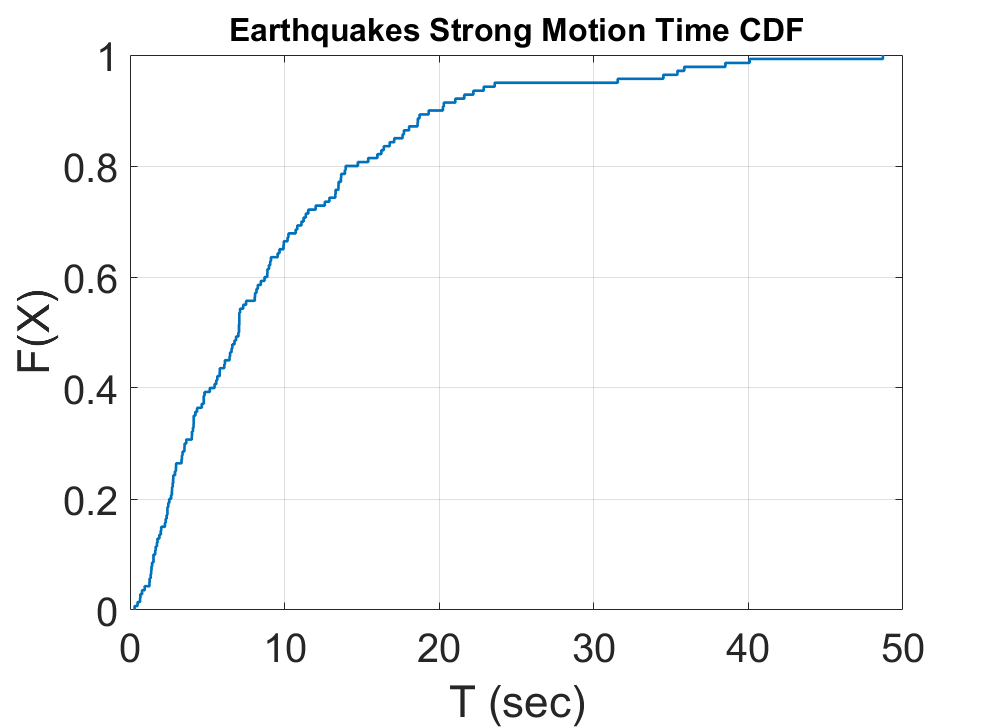}
	\caption{CDF of strong motion duration of 140 horizontal components of earthquake ground motion recorded in California reported in \cite{quakes01}.}
	\label{fig:quakes_CDF}
\end{figure}

Damage prone buildings will suffer damage during the strong shaking part of the seismic event. As mentioned in \cite{quakes01}, there are several definitions for the strong motion duration, which is calculated based on acceleration magnitude or cumulative energy obtained by integrating squared acceleration. In \cite{quakes01}, strong motion duration of 140 earthquake records were evaluated, and figure \ref{fig:quakes_CDF} shows the \ac{CDF} of strong motion duration.

\acrodef{CDF}{cumulative density function}

\subsection{Probability of Classification Error}
\begin{figure}
	\centering
	\includegraphics[width=1.0\linewidth]{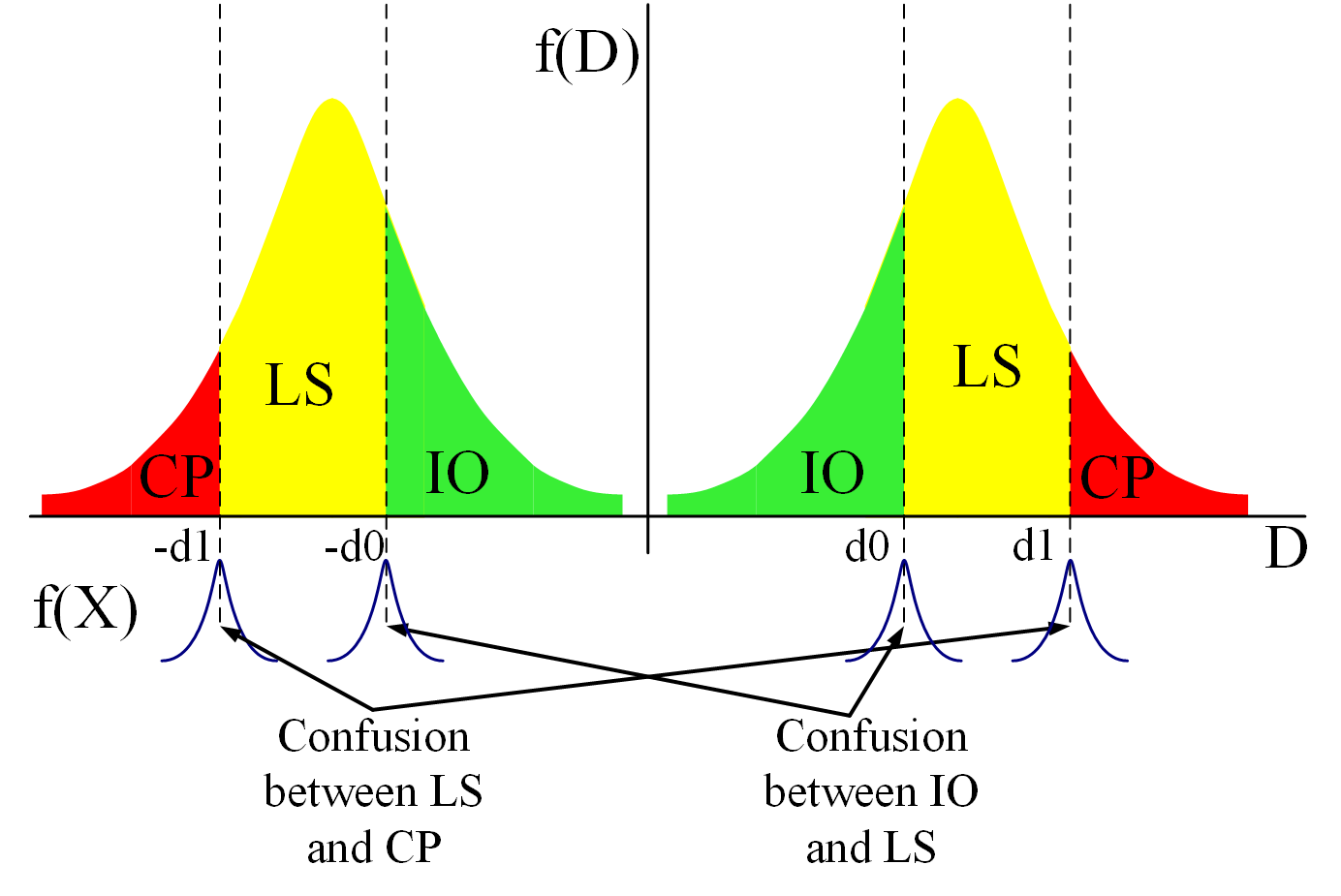}
	\caption{Sketch of peak relative displacement Gaussian distribution.}
	\label{fig:IDR_gaussian}
\end{figure}

As mentioned in section \ref{sec:modeling}, relative displacement measurement  error can be modeled as zero mean Gaussian of variance $\sigma^2_X$, and relative displacement distribution can be modeled as Gaussian of mean $\mu_D$ and variance $\sigma^2_D$ which varies according to the hazard level. As a result, substituting into (\ref{eqn:joint_pdf1}), the joint probability distribution $f_{X,D}(x,d)$ can be expressed by (\ref{eqn:joint_pdf2}).

\begin{align}
f_{X,D}(x,d) =&\mathcal{N}(0,\sigma_X^2)\mathcal{N}(\mu_D,\sigma_D^2) \label{eqn:joint_pdf2}
\end{align}
where $\mathcal{N}(\mu,\sigma^2)$ is a Gaussian distribution with mean $\mu$ and variance $\sigma^2$. Figure \ref{fig:IDR_gaussian} shows a sketch of Gaussian peak relative displacement distribution. Classification boundaries are highlighted, where error is expected to occur.

Substituting (\ref{eqn:joint_pdf2}) in (\ref{eqn:ptrue}) and (\ref{eqn:joint_event}), conditional probabilities can be calculated. For instance, for \ac{IO} buildings, the probability of correct classification is defined as $P(IO|IO)$, whereas probability of error is defined as $P(\overline{IO}|IO)$, i.e. $P(LS \cup CP |IO)$. 
%
Similarly, for \ac{LS} buildings, the probability of correct classification is defined as $P(LS|LS)$, whereas the probability of error is defined as $P(IO \cup CP |LS)$, and with respect to \ac{CP} buildings, the probability of correct classification is defined as $P(CP|CP)$, whereas the probability of error is defined as $P(IO \cup LS |CP)$. The probability of error of the system is calculated by (\ref{eqn:pe}), and will be used later in section \ref{sec:case_study} for sensor selection.

\begin{align}
p_e 	=	&(P(LS|IO) + P(CP|IO))P(IO) +\nonumber\\\
&(P(IO|LS) + P(CP|LS))P(LS) +\nonumber\\
&(P(IO|CP) + P(LS|CP))P(CP) \label{eqn:pe}
\end{align}

\section{Sensor Selection}
\label{sec:case_study}
In section \ref{sec:classification} we showed that the probability of classification error is a function of displacement measurement accuracy, hazard level and strong motion duration. In this section, we demonstrate how a sensor can be selected based on the acceptable probability of error $p_e$ which is calculated by (\ref{eqn:pe}). The probability of classification error is calculated for each of the sensors mentioned in section \ref{sec:noise}.

\begin{figure}
	\centering
	\includegraphics[width=1.1\linewidth]{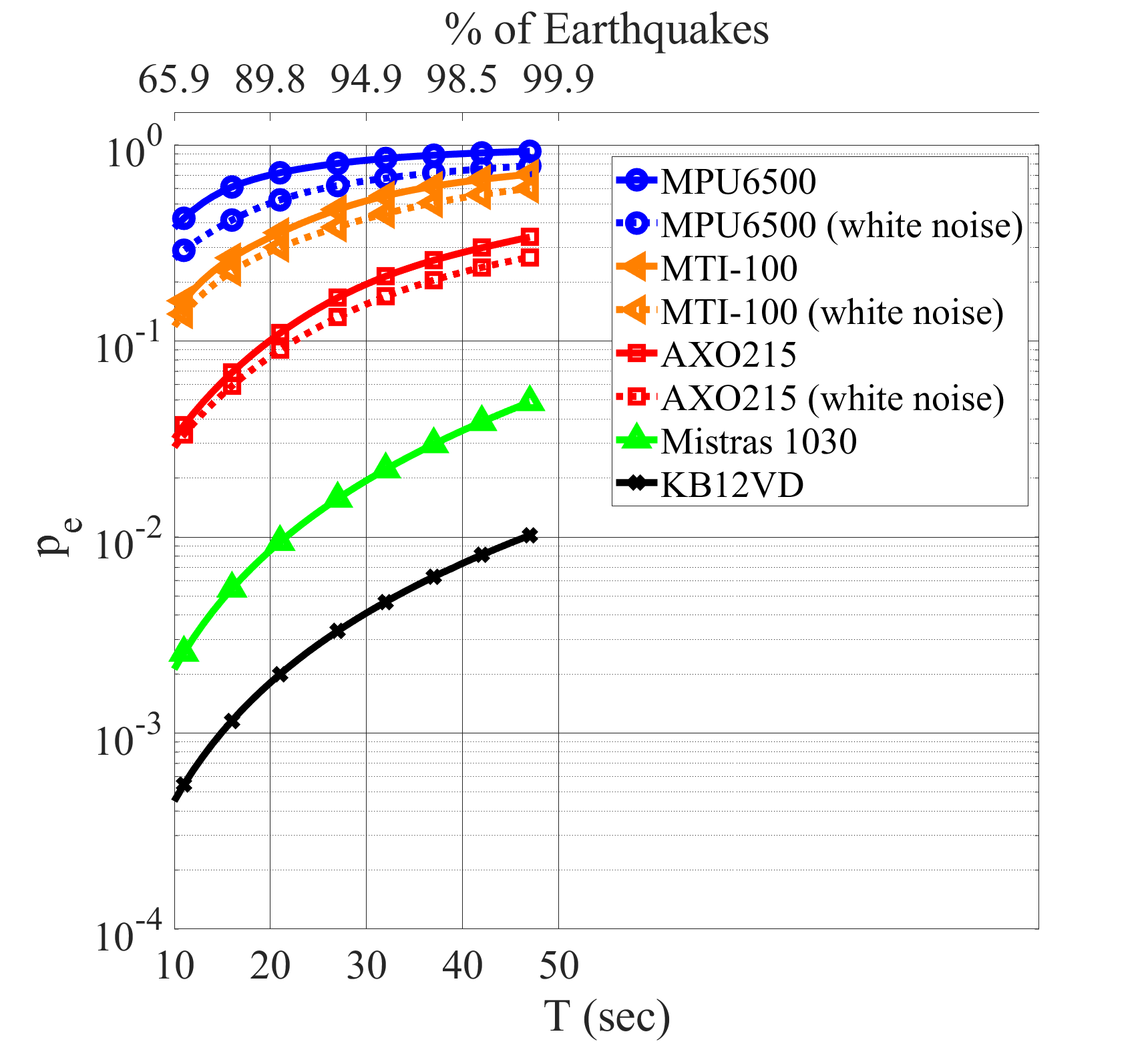}
	\caption{Probability of error in classification $p_e$ versus strong motion duration time $T$ for several sensors, for 50\% in 50 years hazard level. Sensors noise is modeled according to their data sheets except for MPU6500 we used the noise model mentioned in section \ref{sec:noise}.}
	\label{fig:pe_weak}
\end{figure}

\begin{figure}
	\centering
	\includegraphics[width=1.1\linewidth]{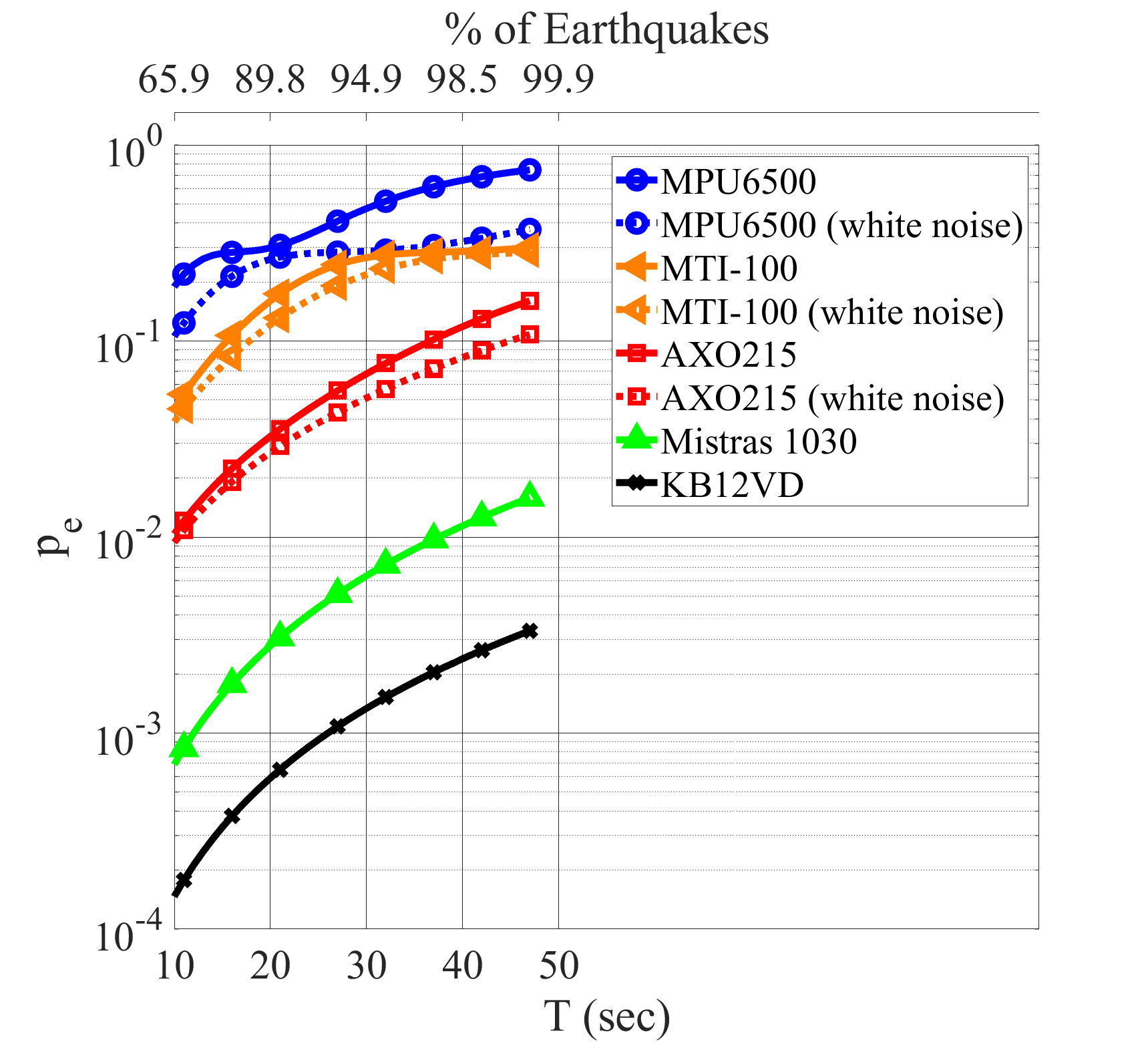}
	\caption{Probability of error in classification $p_e$ versus strong motion duration time $T$ for several sensors, for 10\% in 50 years hazard level.  Sensors noise is modeled according to their data sheets except for MPU6500 we used the noise model mentioned in section \ref{sec:noise}.}
	\label{fig:pe_medium}
\end{figure}

\begin{figure}
	\centering
	\includegraphics[width=1.1\linewidth]{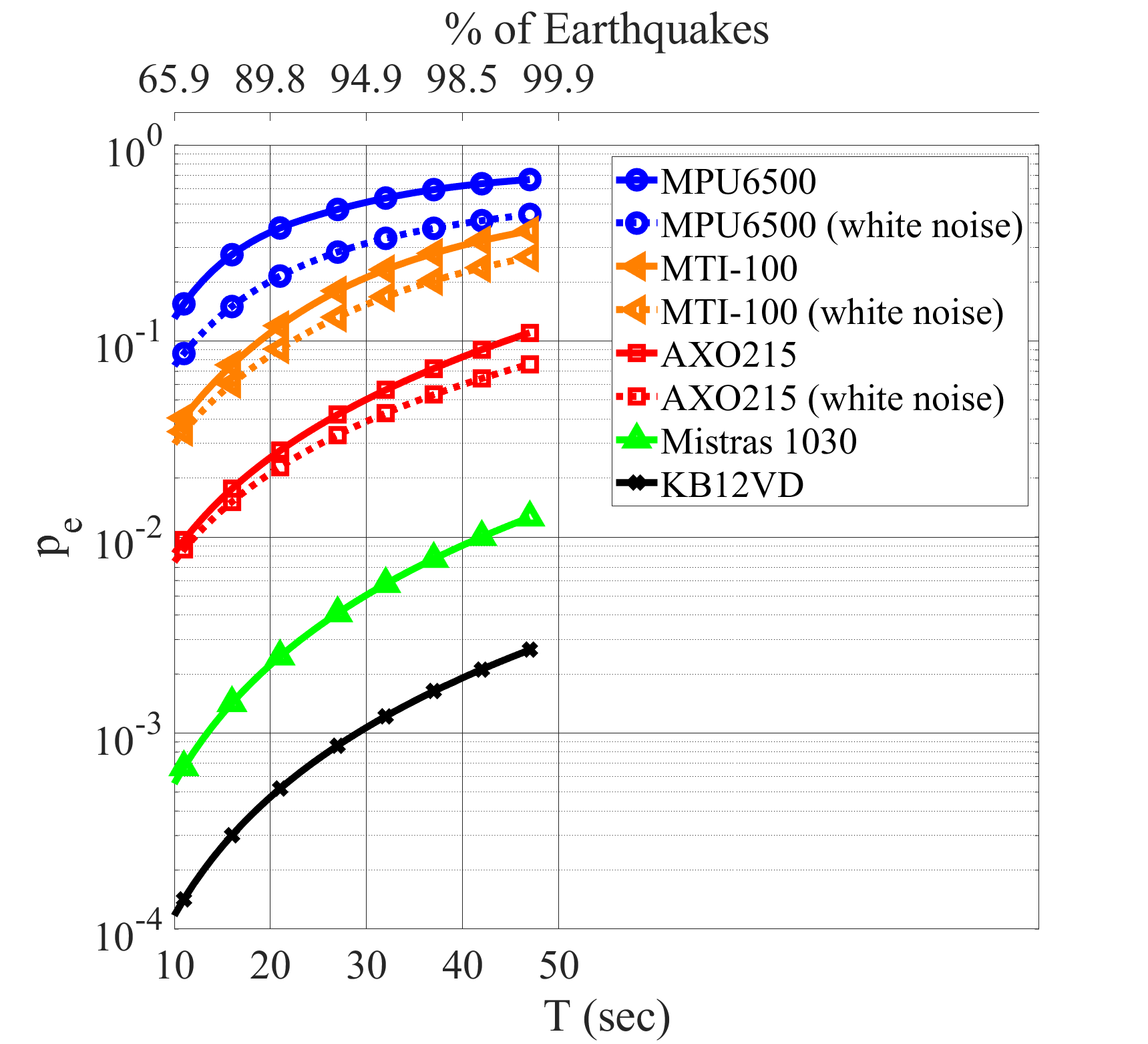}
	\caption{Probability of error in classification $p_e$ versus strong motion duration time $T$ for several sensors, for 2\% in 50 years hazard level. Sensors noise is modeled according to their data sheets except for MPU6500 we used the noise model mentioned in section \ref{sec:noise}.}
	\label{fig:pe_strong}
\end{figure}

Figure \ref{fig:pe_weak} shows the probability of error in buildings classification as a function of strong motion duration in case of 50\% in 50 years hazard level. As expected, it is clear that the high accuracy seismic sensors such as Mistras1030 and KB12VD have very small probability of error, and the probability of error increases as sensor accuracy decreases. In the same figure, we also compare between the simple white noise model, and the more complex model that takes into account other noise components. It is worth noting that for high accuracy sensors, using white noise model results in negligible probability of error which is not plotted in the figure. Hence, only the more complex noise model is plotted for the two high accuracy sensors Mistras 1030 and KB12VD. However, for MTI-100 and AXO215, using only the simple white noise model results in probability of error slightly smaller but comparable to the complex model. With respect to MPU6500, there is a larger discrepancy between white noise model and complex noise model results.   Intuitively, that result was expected, as mentioned in section \ref{sec:noise} by comparing the noise density curves shown earlier in figure \ref{fig:PSDsensors}, it is clear that only MTI-100 and AXO215 noise can be approximated as flat white noise. Similarly, figures \ref{fig:pe_medium} and \ref{fig:pe_strong} show the probability of error in case of 10\% in 50 years and 2\% in 50 years hazard levels respectively. Depending on the acceptable probability of error, the curves presented in figures \ref{fig:pe_weak} to \ref{fig:pe_strong} can be used to evaluate the maximum accepted noise density, hence an appropriate sensor can be selected.

\section{Conclusion}
\label{sec:conclusion}
Monitoring structural health of buildings during and after natural disasters is crucial, and directly impacts public safety. Buildings can be added to an \ac{IoT} network by deploying inertial sensors in civil infrastructure, which  facilitates post disaster identification of structurally unsound buildings. 
In this work, we illustrated how accelerometer sensors can be employed to identify buildings damage state.  We presented a theoretical derivation of a \ac{ZUPT} algorithm that is used to increase displacement measurement accuracy, and consequently increase buildings classification accuracy. The developed algorithm has been validated experimentally using shake table experiments. We investigated the effect of sensors inherent noise on the overall building classification accuracy. The probability of error was calculated as a function of sensor noise density, earthquake duration time and \ac{IDR} distribution. 

\textcolor{black}{
While the focus of this paper is accelerometers, we believe that hybrid systems that combine multiple modalities (e.g. accelerometers + GPS + Camera) will provide enhanced accuracy over a single modality. The trade-offs involved in these systems will be the subject of future work.
}

\bibliographystyle{IEEEtran}
\bibliography{sample}




\end{document}